\documentclass[12pt, letterpaper, oneside]{article}
\usepackage{graphicx}
\usepackage{epstopdf}
\usepackage[body={6.5in, 9in}, right=1in, top=1in]{geometry}
\usepackage{amssymb}
\usepackage{amsmath} 

\newcommand\ee{\end{equation}}
\newcommand\be{\begin{equation}}
\newcommand\eea{\end{eqnarray}}
\newcommand\bea{\begin{eqnarray}}
\newcommand{\sfrac}[2]{{\textstyle\frac{#1}{#2}}}
\newcommand\di{\partial}
\newcommand\mpl{M_{\rm Pl}}

\begin{document}
\def\thefootnote{\fnsymbol{footnote}}

\begin{center}
\Large{\textbf{Galilean Genesis:\\
an alternative to inflation}} \\[0.5cm]
 
\large{Paolo Creminelli$^{\rm a}$, Alberto Nicolis$^{\rm b}$ and Enrico Trincherini$^{\rm c,d}$}
\\[0.5cm]

\small{
\textit{$^{\rm a}$ Abdus Salam International Centre for Theoretical Physics\\ Strada Costiera 11, 34151, Trieste, Italy}}

\vspace{.2cm}

\small{
\textit{$^{\rm b}$ Department of Physics and ISCAP, \\ Columbia University, New York, NY 10027, USA}}

\vspace{.2cm}

\small{
\textit{$^{\rm c}$ SISSA, via Bonomea 265, 34136, Trieste, Italy}}

\vspace{.2cm}

\small{
\textit{$^{\rm d}$ INFN - Sezione di Trieste, 34151 Trieste, Italy}}

\vspace{.2cm}

\end{center}

\vspace{.8cm}

\hrule \vspace{0.3cm}
\noindent \small{\textbf{Abstract}\\
We propose a novel cosmological scenario, in which standard inflation is replaced by an expanding phase with a drastic violation of the Null Energy Condition (NEC): $\dot H \gg H^2$. The model is based on the recently introduced Galileon theories, which allow NEC violating solutions without instabilities. The unperturbed solution describes a Universe that is asymptotically Minkowski in the past, expands with increasing energy density until it exits the regime of validity of the effective field theory and reheats. This solution is a dynamical attractor and the Universe is driven to it, even if it is initially contracting. The study of perturbations of the Galileon field reveals some subtleties, related to the gross violation of the NEC and it shows that adiabatic perturbations are cosmologically irrelevant. The model, however, suggests a new way to produce a scale invariant spectrum of isocurvature perturbations, which can later be converted to adiabatic: the Galileon is forced by symmetry to couple to the other fields as a dilaton; the effective metric it yields on the NEC violating solution is that of de Sitter space, so that all light scalars will automatically acquire a nearly scale-invariant spectrum of perturbations.} 
\vspace{0.3cm}
\noindent
\hrule
\def\thefootnote{\arabic{footnote}}
\setcounter{footnote}{0}

\section{Introduction}

The connection between early cosmology and high energy dates back to the discovery of the expansion of the Universe. Going backwards in time, the Universe contracts and the energy density increases more and more until it eventually reaches Planckian values and General Relativity breaks down. The underlying assumption to this general argument is that the stress-energy tensor $T_{\mu\nu}$ satisfies the NEC, which states that $T_{\mu\nu} k^\mu k^\nu \geq 0$ for every null vector $k^\mu$. For a perfect fluid this is equivalent to the inequality $\rho + p \geq 0$, where $\rho$ and $p$ are respectively the energy density and the pressure. For a Friedmann-Robertson-Walker (FRW) metric this inequality implies that the energy density, and therefore the Hubble parameter $H$, decreases as the Universe expands, as the covariant conservation of the stress-energy tensor reads $\dot \rho = - 3 H (\rho +p)$.  

If a violation of the NEC were possible, then a Pandora's box of non-standard cosmologies would open up and in particular the contraction of the Universe going backwards in time would not necessarily lead to higher and higher energy densities. The realization that energy conditions can be violated, although ``standard" matter satisfies them, has a notable history in cosmology. Indeed the strong energy condition, equivalent for a perfect fluid to the inequality $\rho + 3 p \geq 0$, implies that the expansion of the Universe is always decelerating $\ddot a <0$, which resonates well with the Newtonian intuition. The evidence of the present acceleration, however, strongly indicates that the fluid which now dominates the Universe violates the strong energy condition and the same happens in the past during inflation, the most compelling theory of the early Universe. Given that these two important revolutions in cosmology are based on a violation of the strong energy condition, it is natural to wonder whether we are missing something taking the NEC for granted.
It is worth emphasizing  that the NEC is usually taken for granted not due to lack of imagination, but because of its exceptional robustness---it is especially hard to construct consistent effective field theories that violate it \cite{thomas}. Morevover, perhaps less decisively, the NEC protects standard properties of black-hole theormodynamics.

In this paper we describe a novel cosmological scenario in which the NEC is grossly violated, and the Universe starts from a very low energy state, asymptotic to Minkowski in the far past. As pointed out in \cite{Nicolis:2009qm}, the violation of the NEC is possible in the context of the recently studied Galileon theories \cite{Nicolis:2008in}. For these theories the usual relation between the violation of the NEC and the presence of pathological instabilities \cite{thomas} is avoided, due to the presence of higher derivative interactions. A similar situation happens in the context of Ghost Condensate theories \cite{ArkaniHamed:2003uy,markus}, with the important difference that here we will have a strong violation of the NEC, i.e.~$\dot H \gg H^2$, while in the previous models, only a moderate violation of the NEC is compatible with the stability of the system \cite{markus}.

A general class of Galileon theories, as we will see in Section \ref{background}, gives rise to a very peculiar evolution of the scale factor with the Hubble parameter becoming larger and larger as the Universe expands: $H \propto (-t)^{-3}$ as $t \to 0^-$. This implies that most of the energy is created suddenly, with the scale factor blowing up as $a \sim \exp(1/t^2)$, in a sort of Genesis which (partially) justifies our dubbing of the scenario. As $\rho$ increases, the system will eventually exit the regime of validity of the Galileon effective field theory and here we assume that the energy gets transferred to more conventional degrees of freedom in a reheating process, similarly to what happens in inflation. This background evolution is completely stable and it represents a dynamical attractor. Notably the Universe evolves to this expanding phase even if it is initially contracting, a behaviour which is only possible because of the violation of the NEC. It is remarkable that this scenario in some sense explains why the Universe is now expanding, while we are usually forced to postulate initial conditions with a large positive $H$, which then goes on decaying for the entire evolution.\footnote{Of course our `auto-expanding' solution run backwards in time is also a solution---i.e.~we also have solutions that contract for all times and approach Minkowski space in the future. What we want to stress is that we  are driven towards our expanding solution starting from an unusually large basin of attraction, which includes contracting initial conditions as well. 
The time-reversed solutions we just alluded to start outside this basin.}

The study of perturbations in Section \ref {perturbations} and Appendix \ref{details} shows various peculiarities of this model. One such peculiarity is that the energy density of the background solution vanishes in the limit in which gravity is decoupled, $\rho \propto 1/\mpl^2$. Another peculiar feature is that the leading adiabatic solution does not correspond to the standard $\zeta =$ const.~mode, but to a constant time shift of the unperturbed solution (this will be shown in Newtonian gauge in section \ref{Newtonian}). Such a  mode is going to decay during the standard post-reheating FRW evolution and this will allow us to conclude that the fluctuations of the Galileon field do not give rise to any relevant cosmological perturbations on large scales. Actually in Appendix \ref{squeezing} we will show that the Galileon perturbations are not amenable to any classical interpretation as they experience no relevant squeezing.

Fortunately, another source of scale invariant perturbations is naturally present in our model, as we explain in Section \ref{fake}. The Galileon Lagrangian is invariant under the conformal group SO(4,2) and the time dependent solution breaks it down to SO(4,1), the isometry group of de Sitter space. The only way another field can couple to the Galileon while respecting the conformal symmetry is by treating the Galileon as a dilaton, that is through a fictitious, conformally flat, metric. Therefore all other fields will perceive the Galileon background as a ``fictitious" de Sitter space and their dynamics will be essentially the same as for inflation, even though the Einstein metric at the time when cosmological perturbations are generated is virtually flat. In particular a massless scalar will acquire a scale-invariant spectrum of perturbations. These isocurvature perturbations can then be converted to adiabatic in a variety of ways, exactly as it happens for inflation. This novel mechanism to produce a scale invariant spectrum of perturbations shares some similarities with the attempts to explain the present acceleration through the universal coupling of matter to a scalar field. In both cases, an approximate de Sitter space is realized not in the Einstein metric but in the Jordan one.
However, it is crucial to keep in mind that this ``fake'', Jordan-frame de Sitter space seen by fluctuations, is by no means helping us solve the horizon problem and that the peculiar cosmological history we outlined here and that we are going to describe at length in the paper happens in the {\em Einstein} frame. Indeed our system violates the NEC even in the absence of dynamical gravity \cite{Nicolis:2009qm}; if we do turn on dynamical gravity, coupling it minimally to our system, this NEC-violating stress-energy tensor will generate an Einstein-frame NEC-violating geometry. Related to this, we also notice that the Galileon was originally introduced as a possible explanation of the present acceleration \cite{Nicolis:2008in}, benefiting from its natural implementation of the Vainshtein screening mechanism at short scales. Here we are using it in a completely different context, motivated by its healthy violation of the NEC.

It is important to stress a problem with our model: superluminality. Perturbations around the SO(4,1) invariant background move at the speed of light due the large amount of residual symmetry, and actually gravity corrections to the solution make them slightly subluminal. On the other hand, if we allow for large departure from the background, perturbations around the new solution will move superluminally. In Section \ref{super}, we will discuss this issue and its implications applying to our case the general discussion of \cite{Nicolis:2009qm}. Conclusions are drawn in Section \ref{conclusions}.

\section{The background}\label{background}

Our starting point is the simplest version of the conformal Galileon minimally coupled to gravity: the Lagrangian for the scalar field is just the sum of the kinetic term and the Galilean invariant cubic interaction plus the $(\partial \pi)^4$ term needed to recover conformal invariance \cite{Nicolis:2008in}
\be \label{minimal}
{\cal S}_\pi = \int \! d^4 x \, \sqrt{-g} \bigg[ f^2 e^{2 \pi} (\di \pi)^2 + \frac{f^3}{\Lambda^3} (\di \pi)^2  \Box \pi 
+ \frac{f^3}{2 \Lambda^3} (\di \pi)^4 \bigg] \; ,
\ee
Lorentz indices are contracted with $g_{\mu\nu}$ and the $\Box$ contains a covariant derivative\footnote{We are using the mostly plus signature.}. Notice that the conformal symmetry of the $\pi$ Lagrangian is explicitly broken by the coupling with gravity. 
We could add all Galilean-invariant interactions together with their conformal completions \cite{Nicolis:2008in}, and in fact a fully consistent NEC-violating system obeying all requirements of \cite{Nicolis:2009qm} will have them. However our analysis and results below would not be affected in an essential way, since virtually all our results follow from the symmetry structure of the theory. One important difference is that, with our minimal Lagrangian, the kinetic term has the wrong, ghost-like sign, around the trivial background $\pi=0$, while it is healthy in more general Galilean Lagrangians  \cite{Nicolis:2009qm}. This instability is not relevant for us as we are going to be interested in a different background solution, but it may become important if the system eventually evolves to $\pi = 0$ after reheating.
In the present analysis we stick to the minimal theory (\ref{minimal}), for the simplicity of the computations involved.
We will comment on the effect of adding higher order Galilean terms when relevant. 

The signs have been chosen so that if gravity is decoupled this action has a solution in Minkowski spacetime of the form
\be
\label{pidesitter}
e^{\pi_{{\rm dS}}} = -\frac{1}{H_0 t} \;, \qquad -\infty < t < 0 \; ,
\ee
provided that
\be \label{H0}
H_0^2 = \frac{2 \Lambda^3}{3 f} \; .
\ee
Such a solution spontaneously breaks the conformal group $SO(4,2)$ down to the de Sitter group $SO(4,1)$.
More importantly for our purposes, this ``de Sitter" field configuration violates the NEC but has no instabilities  \cite{Nicolis:2009qm}. The $\pi$ stress-energy tensor can be easily computed from the action (\ref{minimal}) as $T_{\mu\nu} = -\frac{2}{\sqrt{-g}}\frac{\delta S_\pi}{\delta g_{\mu\nu}}$,  
\bea \label{Tmn}
T_{\mu\nu}  
& = & - f^2 e^{2 \pi}\big[ 2\di_\mu \pi \di_\nu \pi -  g_{\mu\nu} (\di \pi)^2 
		\big] \nonumber\\
&-& \frac{f^3}{\Lambda^3} \big[ 2 \, \di_\mu \pi \di_\nu \pi  \Box \pi - \big(\di_\mu \pi  \, \di_\nu ( \di \pi)^2 + \di_\nu \pi \,  \di_\mu ( \di \pi)^2 \big)
+  g_{\mu\nu} \, \di_\alpha \pi \,  \di^\alpha ( \di \pi)^2
\big] \nonumber  \\
&-& \frac{f^3}{2 \Lambda^3} \big[ 4 ( \di \pi)^2 \di_\mu \pi \di_\nu \pi - g_{\mu\nu} ( \di \pi)^4 
		\big] \; .
\eea
By plugging the solution (\ref{pidesitter}) into this expression with $g_{\mu\nu}= \eta_{\mu\nu}$ we find that it has vanishing energy density---this is a consequence of scale-invariance which is left unbroken by the background \cite{Nicolis:2009qm}---and negative pressure $\propto - 1/t^4$.

We defer a thorough stability analysis for this solution until the next section. For the moment, let us consider the dynamics of {\em homogeneous} perturbations  $\delta \pi(t)$. From the Lagrangian (\ref{minimal}) we immediately get the equation for $\delta \pi$ in the linear regime:
\be\label{perturbdec}
\delta \ddot \pi - \frac{2}{t} \delta \dot \pi -\frac{4}{t^2} \delta \pi  = 0 \;;
\ee
the two independent solutions are $\delta \pi \sim 1/t$ and  $\delta \pi \sim t^4$. The latter decays away for $t \to 0^-$ and is thus no source of worry. The former blows up at late times, but in fact it just describes the same background solution we are interested in, slightly shifted in time \cite{Nicolis:2009qm}:
\be
\pi_{\rm dS} (t + \epsilon) \simeq \pi_{\rm dS} (t) + \dot \pi_{\rm dS} (t ) \cdot \epsilon = \pi_{\rm dS} (t) - \frac \epsilon t \; .
\ee
We conclude that a generic homogeneous initial condition that corresponds to a small departure from $\pi_{\rm dS}$ will be diluted away at late times.  
   
Now let's reintroduce the coupling to gravity: the presence of a non-zero stress-energy tensor will source a gravitational field and $\pi=\pi_{\rm dS}$ with flat metric will no longer be a solution. However, since also the pressure vanishes at early times ($t \rightarrow -\infty $) the gravity-free solution is recovered in this limit,
while as time goes on corrections to this asymptotic behavior will become larger and larger. As we are interested in cosmological solutions of the form
\be
ds^2 = -dt^2 + a^2(t) d \vec x^2 \, , \qquad \qquad \pi=\pi_0(t) 
\ee
we have just to solve Friedmann's equations for the Hubble rate $H$,
\bea
H^2 & = & \sfrac{8 \pi}{3} G \, \rho  \label{F1}\\
\dot H  & = & -4\pi G (\rho + p) \, \label{F2} 
\eea
and the $\pi$ e.o.m. will be automatically satisfied.   

From eq.~(\ref{Tmn}) we get the energy density and the pressure,
\bea
\rho & = & - f^2 \left[ e^{2\pi} \dot \pi^2 - \frac{1}{H_0^2} \big( \dot \pi^4 + 4  H \dot \pi^3 \big) \right] \label{rho} \\
p & = & - f^2 \left[ e^{2\pi} \dot \pi^2 - \frac{1}{3 H_0^2} \big( \dot \pi^4 - \sfrac43 \sfrac{d}{d t} \dot \pi^3 \big) \right] \;,
\eea
where we used eq.~(\ref{H0}) to write $\Lambda^3$ in terms of $H_0$.
We cannot solve Friedmann's equations analytically. Still, we can compute the asymptotic behaviors of the solution at early and late times. As we said, at early times the solution is approximately the unperturbed one $\pi=\pi_{\rm dS}$ and $H=0$, with small corrections proportional to $G= \frac{1}{8 \pi M_{\rm Pl}^2}$, which we can calculate perturbatively. Since $\rho \sim {\cal O}(G) \ll p = p_{\rm dS} + {\cal O}(G)$, the leading contribution to eq.~(\ref{F2}) is $\dot H \simeq -4\pi G \, p_{\rm dS} $ and this expression can be integrated to find the Hubble rate at early times:
\be\label{earlytimeH}
  H \simeq - \frac{1}{3} \frac{f^2}{M_{\rm Pl}^2} \cdot \frac{1}{H_0^2 t^3} \quad {\rm for }\;  t \to -\infty \; .
\ee
This result has a number of unusual properties: {\em i)} we can add to the value of $H$ an arbitrary integration constant and still have a solution of eq.~(\ref{F2}); we will discuss this possibility shortly, while for the moment we set the constant to zero; {\em ii)} because $\rho \rightarrow 0$ in the limit $M_{\rm Pl} \rightarrow \infty$, the Hubble rate is proportional to $1/M^2_{\rm Pl}$, unlike for standard cosmological scenarios where it scales like $1/M_{\rm Pl}$; {\em iii)} $H$ increases with time as a consequence of the NEC violation, with a rate $\dot H \gg H^2$.
 
Having computed the value of $H$ we can plug it into (\ref{F1}), or equivalently into the scalar equation of motion, and extract the ${\cal O}(G)$ correction to $\pi_{\rm dS}$ 
\be\label{earlytimepi}
t \to -\infty \qquad  \pi_0 \simeq \pi_{\rm dS} - \frac{1}{2} \frac{f^2}{M_{\rm Pl}^2} \cdot \frac{1}{H_0^2 t^2} 
\ee
(we are choosing $\pi_0 \to \pi_{\rm dS}$ for $t \to -\infty$ as initial condition, as required by consistency of the approximations we have adopted so far.)
At late times, for $t^2 \lesssim \frac{f^2}{M_{\rm Pl}^2} \frac{1}{H_0^2}$ the above approximation breaks down. Numerically integrating the Friedmann's equations shows that both $\pi$ and $H$ diverge at some positive $t_0 \sim H_0^{-1} \frac{f}{M_{\rm Pl}}$. Then, assuming they diverge like some powers of $(t_0 -t)$, we get their asymptotic behaviors:
\bea
\label{polephase}
t \to t_0 \quad && e^{\pi_0} \simeq \frac{8}{\sqrt{3}} \frac{f}{M_{\rm Pl}} \cdot \frac{1}{H_0^2 (t_0 - t)^2} \\ \label{polephase2}
&&  H \simeq  \frac{16}{3}  \frac{f^2}{M_{\rm Pl}^2} \cdot \frac{1}{H_0^2 (t_0 - t)^3} \;,
\eea
which indeed match the actual numerical solutions. This gives the peculiar evolution of the scale factor
\be
a(t) \sim \exp{\left[\frac{8 f^2}{3 H_0^2 M_{\rm Pl}^2} \frac{1}{(t_0-t)^2}\right]}  \; . 
\ee   

\begin{figure}[!!!t]
\begin{center}
\includegraphics[scale=0.45]{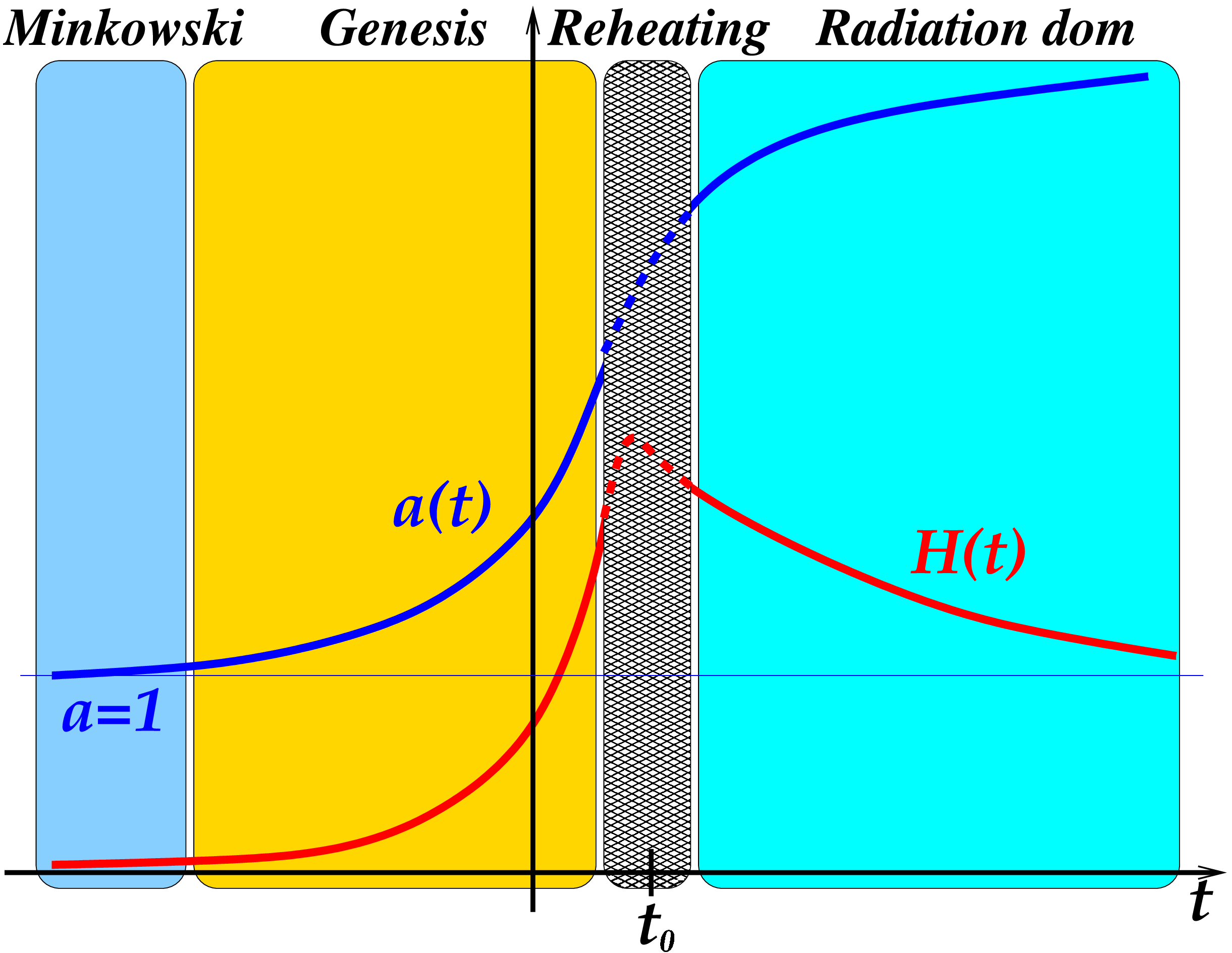}  
\end{center}
\caption{\small {\em The cosmological evolution in our model.}}
\label{fig:cartoon}
\end{figure}

We have then the following scenario, represented in figure \ref{fig:cartoon}. The Universe starts at $t \to -\infty$ in a quiescent state, with flat metric and $\pi = \pi_{\rm dS}(t)$. Asymptotically in the past this configuration has zero stress-energy tensor, and so it is a solution. Then, as time goes by, a negative pressure arises, $p \sim -1/ t^4$, which makes the Universe start expanding and the energy density grow. The actual solution then departs from the original ``de Sitter'' configuration. $H$, $\rho$, and $p$ grow more and more, until $\pi$ becomes strongly coupled. At that moment the effective theory of $\pi$ breaks down, and we cannot predict what happens next. We can imagine that at that point, most of the available energy gets converted into radiation, the Universe reheats, and the standard radiation-dominated era takes over.  In this era the Galileon $\pi$ may evolve to the $\pi =$ const.~solution, or cease to be a good degree of freedom.

The strong-coupling scale of the theory ``runs'' with $\phi \equiv H_0 e^\pi$, and it is \cite{Nicolis:2009qm} 
\be \label{strongcoupling}
\Lambda_{\rm strong} \sim \frac{\phi}{g^{1/3}} \; , \qquad g \equiv H_0/f \; .
\ee
Notice that this estimate does not take into account gravity and the associated explicit breaking of conformal symmetry. It may thus get modified when dynamical gravity is included. Our effective theory breaks down---at the latest---when typical energies become larger than $\Lambda_{\rm strong}$. A good measure of a cosmological solution's typical energy is the freeze-out frequency for fluctuations: if fluctuations are strongly coupled at freeze-out, the background solution is hardly consistent.
In the phase where things are blowing up, eqs.~(\ref{polephase}, \ref{polephase2}), freeze-out happens at frequencies of order $H$. The highest Hubble rate we can get before strong-coupling/reheating thus comes from equating eq.~(\ref{polephase2}) with the strong-coupling scale (\ref{strongcoupling}). We get an impressive 
\be
H_{\rm max} \sim M_{\rm Pl} \; ,
\ee
at which GR breaks down anyway. Because of this, we don't want to go this high in $H$, so we will assume that our effective theory breaks down before becoming strongly coupled, which is also very much in line with the arguments presented in \cite{Nicolis:2009qm}.
Notice that all other measures of the solution's typical energy---for instance $\dot \phi/\phi$ etc.---are smaller that the one we have adopted, and are thus less constraining.
In conclusion, the reheating temperature is essentially unconstrained in our model.

At this point one may ask whether the background discussed so far is an attractor. As in the previous discussion in the absence of gravity, we can study the homogenous perturbations $ \pi (t) =  \pi_0(t) + \delta \pi(t) = \pi_{\rm dS}(t) - \frac{1}{2}\frac{f^2}{M^2_{\rm Pl}} \frac{1}{(H_0 t)^2} +  \delta \pi(t) $. To expand linearly in $\delta \pi$ the only conditions needed are $\delta \pi \ll \pi_0$ and $\delta \pi \ll 1$. However even in the linearized approximation there are two possible regimes, depending on whether the perturbations are also smaller than the gravitational corrections suppressed by $1/M^2_{\rm Pl}$ or not. 

Let's start from the scalar equation of motion:
\be\label{pieom}
 e^{2\pi} (\ddot \pi + \dot \pi^2) - \frac{2}{H_0^2} \dot \pi^2 \ddot \pi = -3 e^{2 \pi} H \dot \pi +\frac{1}{H_0^2}\big[ 4 H \dot \pi \ddot \pi + 2 \dot \pi^2 (3 H^2 + \dot H)+ 2H \dot \pi^3 \big] \; .
\ee
Keeping $\delta\pi$ fixed and going to early times we can neglect gravity corrections and the linearized equation of motion for $\delta \pi$ reduces to the one studied before, eq. (\ref{perturbdec}), with the two solutions $\{1/t; \, t^4 \}$.
Notice another unfamiliar feature of this background: since $\rho_{\rm dS}=0$ and we are in a regime where gravitational contributions are small, linear perturbations to the scalar give a contribution to $H$ larger than that of the background.

In the opposite regime $\delta \pi \ll \frac{f^2}{M^2_{\rm Pl}} \frac{1}{(H_0 t)^2}$ terms proportional to $H$, $\dot H$ on the right-hand side of (\ref{pieom}) will give a contribution to the linear equation for the pertubations. Since $H^2$, contrary to the previous case, is now dominated by the background solution we can expand in equation (\ref{F1}) $\delta H^2 = 2H \delta H$ together with $\pi=\pi_0 + \delta \pi$ to find $\delta H = - \delta \pi/ t - \delta \dot \pi$. Using this expression in the RHS of (\ref{pieom}) gives the perturbations' equation in this regime:
\be\label{perturb2}
\delta \ddot \pi + \frac{\delta \dot \pi}{t} - \frac{\delta \pi}{t^2} =0 \; .
\ee      
The two solutions are $\{ 1/t; \, t \}$; we have again the shift in time and a decaying solution that implies convergence to the attractor.

We can now study the most general solution for the Hubble rate. Suppose we start with an initial perturbation $\delta \pi = A t^4$; this gives a constant contribution to the energy density $\rho= -10 A f^2/H_0^2$, which corresponds to the integration constant for $H$ we alluded to below eq.~(\ref{earlytimeH}).
Since this can be positive or negative, a generic initial condition can produce a background that starts in expansion or even in contraction.
 
No matter which sign we choose for the initial condition, the perturbation decays as time goes on and we are eventually driven to the original expanding background.
Notice that as we approach the unperturbed solution, $\delta\pi$ will move from the first regime---where it dominates the energy density and has $\delta\pi \propto t^4$ as solution---to the second one, where $\delta\pi$ is smaller than the perturbation induced by gravity. In the intermediate regime we have no analytic control and one may be worried that we do not recover the unperturbed solution eventually.  However we know that the equation $\dot H \simeq -4 \pi G \, p_{\rm dS}$ is always a good approximation since $\rho \ll p$. It tells us that $H$ follows the evolution discussed above up to small corrections,      
even in the transition between the two limiting regimes for $\delta \pi$, where we don't have an explicit solution for homogenous perturbations.   

In conclusion the NEC violating solution is an attractor for general homogeneous initial conditions close to the de Sitter solution: $\delta\pi \ll \pi_{\rm dS}$.

\section{\label{perturbations}Scalar perturbations}

Let us now move away from homogeneity and discuss scalar perturbations. To begin with, let us first assume again that gravity is decoupled, $\mpl \to \infty$.
Since the solution (\ref{pidesitter}) spontaneously breaks the conformal group $SO(4,2)$ down to the de Sitter one $SO(4,1)$, the Lagrangian for small perturbations will be invariant under the de Sitter symmetries, whereas the broken symmetries will be non-linearly realized.
In particular the fluctuation $\xi(x)$ defined via $\pi(x) = \pi_{\rm dS}(t + \xi(x))$ is the Goldstone boson associated with the spontaneously broken time-translational invariance $t \to t+ \epsilon$, which is now  realized  non-linearly as $\xi \to \xi + \epsilon$.  
Indeed from eq.~(\ref{minimal}) we get the quadratic Lagrangian for $\xi$
\be
\label{phiaction}
{\cal L}_{\xi} = - \frac{f^2}{H_0^2} \frac{1}{t^4} (\partial \xi)^2  \; ,
\ee
which is manifestly shift-invariant.
The kinetic energy is positive, thus ensuring stability  for the background solution  against short-wavelength perturbations. For long-wavelength ones instead we get back to eq.~\eqref{perturbdec}, 
now written in terms of $\xi \,$:
\be
\ddot \xi_k - \frac4t \,  \dot \xi_k = 0 \;, \qquad k \ll 1/t \; .
\ee
The  solutions are
\be
\xi_k \sim t^5 , \: {\rm const}  \; .
\ee
The constant one dominates at late times and simply describes---now manifestly---the original background solution slightly translated in time. We thus conclude that, in the absence of gravity, the solution $\pi_{\rm dS}$ is an attractor also for initial perturbations with non-vanishing gradients.

We now turn on gravity and in general we expect that the dynamics of scalar perturbations at large distances will be modified by their mixing with the scalar sector of $g_{\mu\nu}$.  
Let's see how this works explicitly. 
Suppose we have the background solution
$\pi_0(t)$, $H(t)$. Then, if we consider small fluctuations, it is particularly convenient to work in `unitary gauge':
\be
\pi (\vec x, t) = \pi_0 (t) \; .
\ee
This fixes time-diff invariance: we are defining the equal-time surfaces as the equal-$\pi$ ones, and the pace of time as that of the unperturbed solution. In this case there are no fluctuations in $\pi$, and the scalar fluctuation is in the metric tensor. We will fix the space diffs later.

Following \cite{maldacena} and \cite{markus}, we use ADM variables for the metric: the induced 3D metric $g_{ij}$, the lapse $N \equiv 1/\sqrt{-g^{00}}$, and the shift $N_i \equiv g_{0i}$. It is straightforward (see Appendix \ref{details}) to write the full action, the Einstein-Hilbert one plus the Galileon part, using these variables:
\bea
S  & = & S_g + S_\pi \nonumber \\
S_g & = & \sfrac{1}{2} \mpl^2 \int \! d^4 x \, \sqrt{g_3} N \big[   R_3 + \big( K_{ij}K^{ij} - K^i {}_i {}^2 \big) \big]  \label{Sg} \\
S_\pi & = &  f^2 \int \! d^4 x \, \sqrt{g_3} N \bigg[ 
-  e^{2 \pi_0}\dot \pi_0^2  \,  \frac{1}{N^2} + \frac{4\dot \pi_0^3}{9 H_0^2}   \, \frac{1}{N^3}  K^i {}_i  + \frac{\dot \pi_0^4}{3 H_0^2}   \, \frac{1}{N^4} 
\bigg]   \;, \label{piADM}
\eea
where and henceforth spatial indices are raised and lowered via the spatial metric $g_{ij}$, and we have used the extrinsic curvature of costant-$t$ hypersurfaces 
\be
K_{ij} \equiv \frac{1}{2N} \big[ \di_t g_{ij} - \nabla_i N_j - \nabla_j N_i \big] \; .
\ee

In fact the structure of the action is largely constrained by symmetry considerations \cite{markus}. The background solution spontaneously breaks time translations (diffs) as well as Lorentz boosts. This is made explicit by working with ADM variables, and, in unitary gauge, by allowing for explicit functions of time in the action. Then, we just have to write down all possible operators compatible with the residual symmetries, namely time- and space-dependent spatial diffs, $x^i \to x^i + \xi^i (t, \vec x)$, and 3D rotations.
The generic Lagrangian for matter ($\pi$, in our case) then is \cite{markus}
\bea
S_\pi & = & \int \! d^4 x  \, \sqrt{g_3}N  \nonumber
\left[ - \frac1{8\pi G} \, \dot H \frac{1}{N^2} - \frac1{8\pi G} (3 H^2 + \dot H) \right. \\
& + & \left. \frac12 M^4(t) \, (\delta N)^2 - \hat M^3(t) \,  \delta E^i {}_i  \delta N 
+ \dots \right] \label{ADMaction}\; .
\eea
The terms in the first line are the only `tadpoles' there are: they start linear in the metric perturbations, thus yielding a non-trivial stress energy tensor on the background solution. As a result their coefficients are uniquely determined in terms of the background $H(t)$ by the Friedmann equations. The terms in the second line start quadratic in the fluctuations, and their coefficients are unconstrained. $\delta N$ is obviously the fluctuation in $N$, $N = 1 + \delta N$, whereas the tensor $\delta E_{ij}$ is, apart from an extra factor of $N$, the fluctuation in the extrinsic curvature of constant-$t$ surfaces,
\be
E_{ij} \equiv N K_{ij} \;, \qquad \delta E_{ij} \equiv E_{ij} - a^2 H g_{ij} \; .
\ee
Finally, the dots stand for higher-derivative terms---which in our case vanish, because of the magic properties of our conformally invariant Lagrangian---and for interaction terms, cubic and higher in the metric fluctuations---which we are not interested in. At the quadratic, two-derivative level the action (\ref{ADMaction}) is all we need. 

Our action \eqref{piADM} can indeed be recast in the form \eqref{ADMaction} (see Appendix \ref{details}) with
\be \label{M4}
M^4(t) = \frac43 \frac{f^2}{H_0^2}\left(2 \dot\pi_0^4 + \dot\pi_0^2 \ddot\pi_0 + 9 H \dot\pi_0^3\right) \; ,
\qquad \hat M^3(t) = \frac43 \frac{f^2}{H_0^2} \dot\pi_0^3 \; .
\ee
In the following, however,  we keep the analysis as general as possible,  because then we can 
apply it to other conformally invariant Lagrangians as well.
However from the expression for $\hat M^3$ above we see why we can violate the `theorem' of ref.~\cite{markus}: there it was assumed that the rate at which the  Lagrangian coefficients---in particular $\hat M^3(t)$---vary with time is at most of order $H$. Here however at early times $H \sim f^2/(\mpl^2 H_0^2 t^3)$, whereas 
$(1/\hat M^3) \partial_t \hat M^3 \sim 1/t$, which is much larger than $H$. Thus our example does not satisfy the hypotheses of the theorem. On the other hand, at late times the rate of $\hat M^3(t)$ is slower than the Hubble rate. However at late times $\dot H \ll H^2$, which, according to ref.~\cite{markus} is compatible with a ghost-free violation of the NEC. In conclusion: there is no contradiction with our NEC-violating system being free of instabilities throughout.

Before proceeding, it is time to comment on what changes if we allow for a more general conformal Galilean Lagrangian to start with. The NEC violating background solution will be the same \cite{Nicolis:2009qm}, apart from numerical factors and this fixes the first line of \eqref{ADMaction}. Galilean theories give rise to equation of motion containing at most two derivatives on each field \cite{Nicolis:2008in}, so that at the quadratic level the operators $(\delta N)^2$ and $\delta E^i_i \delta N$ are the only possible ones: all the others would give rise to equation of motion with more than two derivatives on $\xi$ \cite{markus}. With a proper choice of the coefficients of the Galilean Lagrangian the fluctuations around the NEC violating solution are healthy, as in our example eq.~\eqref{minimal}, and it is also possible, at the same time, to have stable perturbations around the $\pi = 0$ background  \cite{Nicolis:2009qm}, i.e.~to flip the worrisome sign of the first term of eq.~\eqref{minimal}. Also the time dependence of  eq.~\eqref{M4} will remain the same in the first phase $|t| \gg t_0$ for a general Galileon theory, so that all our conclusions can be straightforwardly applied to the more general case as well.

We finally move to compute the quadratic action for the propagating scalar mode. This was already done in ref.~\cite{markus} for the Lagrangian (\ref{ADMaction}), but only in the $M^4, \hat M^3 = {\rm const}$ case, which as we argued is quite different from ours.
Following Maldacena \cite{maldacena}, the spatial diffs can be fixed for instance by imposing
\be \label{zetagauge}
g_{ij} = a^2(t) \big[ (1+2 \zeta ) \delta_{ij} + \gamma_{ij} \big] \;, \qquad
\di_i \gamma_{ij} = 0\; , \quad \gamma_{ii} = 0 \; .
\ee
The transverse traceless matrix $\gamma_{ij}$ corresponds to tensor modes, which we will discuss below. For the moment we can consistently set $\gamma_{ij} = 0$, since 3D rotations are left unbroken by the background solution, thus preventing any mixing between scalar and tensor modes at the quadratic level, ${\bf 2} \otimes {\bf 0} \not\supset  {\bf 0}$. $\zeta$ parametrizes the only scalar propagating d.o.f. As we will now see, the remaining metric components, $g_{00}$ and $g_{0j}$, can be expressed as functions of $\zeta$ through the constraint equations. 
These are the variations of the full action $S = S_g + S_\pi$ with respect to $N$ and $N^j$ (see their explicit form in the Appendix \ref{details}). At zeroth-order in the fluctuations, the constraints are solved by the background solution. In particular, the Hamiltonian constraint reduces to Friedmann equation, whereas the momentum constraint is trivial. To get the quadratic Lagrangian for $\zeta$, we need to solve the constraints at first order in the perturbations $\zeta$, $\delta N$, and $N^j$. Defining $N^j = \di_j \beta$
(we can set to zero the transverse vector piece in $N^j$, for the same reason as for the tensor modes), we get
\bea
\label{deltaNzeta}
\delta N & = & \sfrac{2 \mpl^2}{2\mpl^2H - \hat M^3} \, \dot \zeta \\ \label{psizeta}
\nabla^2 \beta & = &  - \sfrac{2 \mpl^2}{2\mpl^2 H - \hat M^3} \, \sfrac1{a^2} \nabla^2 \zeta
+ \sfrac{- 4\mpl^4 \dot H - 12 \mpl^2 H \hat M^3 + 3 \hat M ^6 +  2 \mpl^2  M^4 }{(2 \mpl^2H - \hat M^3 )^2} \, \dot \zeta \;.
\eea
The quadratic action for $\zeta$ is then obtained by plugging these back into the original action, eq.~(\ref{Sg}) plus eq.~(\ref{ADMaction}). After some integrations by parts we get
\be \label{zeta_action}
S_\zeta = \int \! d^4x \, a^3 \left[ A(t) \, \dot \zeta^2 - 
B(t) \, \sfrac1{a^2} \big( \vec \nabla \zeta \big)^2 \right] \; ,
\ee
where
\bea
A(t) & = & \frac{\mpl^2 \big( - 4\mpl^4 \,  \dot H  - 12 \mpl^2 \, H \hat M^3 + 3  \hat M ^6 + 2 \mpl^2 \,  M^4 \big)}{\big(2 \mpl^2 H -  \hat M^3 \big)^2} 
\label{A(t)}\\
B(t) & = & \frac{\mpl^2 \big( - 4\mpl^4 \,  \dot H  + 2 \mpl^2 \, H \hat M^3 -  \hat M ^6 + 2 \mpl^2 \, \di_t  \hat M^3 \big)}{\big(2 \mpl^2 H -  \hat M^3 \big)^2} 
\label{B(t)}
\; .
\eea
As a check, notice that for a cosmology driven by a minimally coupled scalar with a standard kinetic term and non-derivative interactions, we have $\hat M^3 = M^4 = 0$. This implies $A(t) = B(t) = - \mpl^2 \, (\dot H/ H^2)$, which is Maldacena's result \cite{maldacena}.

At early times---or equivalently at leading order in $1/\mpl^2$---we have (see eqs.~(\ref{earlytimeH}) and (\ref{M4}))
\be
\label{expearly}
H \simeq - \frac13 \frac{f^2}{\mpl^2} \cdot \frac{1}{H_0^2 t^3} \;, \quad M^4 \simeq \frac{4 f^2}{H_0^2} \frac1{t^4} \; , \quad \hat M^3 \simeq - \frac{4 f^2}{3 H_0^2} \frac1{t^3} \; , \quad \qquad  |t | \gg \frac{f}{\mpl} H_0^{-1} \; , 
\ee
so that $A(t)$ and $B(t)$ above reduce to
\be
A(t) = B(t) = \frac{9 \mpl^4 H_0^2}{f^2} \, t^2 \; .
\ee
Also, given the smallness of $H$, the scale factor can be approximated as constant, $a(t) = 1 + {\cal O}(1/t^2)$. Therefore at early times we have
\be
\label{zetaaction}
S_\zeta = \frac{9 \mpl^4}{f^2} \int \! d^4x \, (H_0 t)^2 \left[ \dot \zeta^2 - 
\big( \vec \nabla \zeta \big)^2 \right] \; .
\ee

It is easy to deduce the spectrum of $\zeta$ directly from this action. The action is invariant under
\be
t, \vec x \to \lambda t, \lambda \vec x  \qquad \zeta \to \frac{1}{\lambda^2} \zeta \;;
\ee
as a result, the equal-time 2-point function of $\zeta$ must have the general form
\be \label{generic_form}
\langle \zeta (t, 0) \zeta(t, \vec x) \rangle = \frac{f^2} {\mpl^4 H_0^2} \frac{1}{|\vec x|^4} F\big( |\vec x| / t \big) \; ,
\ee
where $F$ is a generic function, with no additional dependence on the model parameters (the normalization prefactor comes from the overall constant appearing in the action (\ref{zetaaction}).) Notice that it is crucial that we look at symmetries of the action rather than simply at symmetries of the equation of motion (under which the action might change by an overall multiplicative constant). Indeed, if we want the $n$-point function to have the same transformation properties as $\zeta^n$, the vacuum state has to be invariant under the symmetries considered. This is the case, barring spontaneous breaking, if the action is invariant. 

At short distances, $|\vec x| \ll t$, we have to recover the standard Minkowski 2-point function for a (non-canonically normalized) massless field:
\be
F\big( |\vec x| / t \big) \sim \frac{ |\vec x|^2}{t^2} \;, \quad  |\vec x| \ll |t| \; .
\ee
To get the behavior of $F$ at large distances, we use the fact that the quantum $\zeta$ solves the classical equations of motion. For long wavelengths, or equivalently late times, we have the two behaviours
\be
\zeta \sim {\rm const}\;,\;\frac1t \;.
\ee
The second term dominates so that $\langle \zeta \zeta \rangle \sim 1/t^2$ also at late times.  This implies\footnote{Using the same method for inflation, one would start with the action in conformal time $\eta$ of the form \cite{maldacena}
\be
\label{inflactionconf}
S= \mpl^2 \epsilon \int d^4 x \;\frac{1}{H^2 \eta^2} \left[\zeta'^2-(\nabla\zeta)^2\right] \;,
\ee
where $H$ and the slow-roll parameter $\epsilon$  can be taken as constants at leading order in slow-roll. This action is invariant under 
\be
\eta, \vec x \to \lambda \eta, \lambda \vec x  \qquad \zeta \to \zeta \;,
\ee
which implies
\be \label{generic_form_infl}
\langle \zeta (\eta, 0) \zeta(\eta, \vec x) \rangle = \frac{H^2} {\mpl^2 \epsilon} F\big( |\vec x| / \eta \big) \;,
\ee
with $F \sim \eta^2/|\vec x|^2$ for $|\vec x| \ll |\eta|$ to reproduce the Minkowski result and $F \sim $ const for $|\vec x| \gg |\eta|$ to reproduce the time evolution $\zeta \sim$ const one deduces from the equation of motion. In Fourier space this gives the celebrated $1/k^3$ spectrum.
}
\be \label{2pf}
\langle \zeta (t, 0) \zeta(t, \vec x) \rangle \propto \frac{f^2} {\mpl^4 H_0^2} \frac{1}{|\vec x|^2 t^2}   \; ,
\ee
which is a very blue spectrum, going as $k^{-1}$ in Fourier space. Indeed the standard calculation in terms of modes (see Appendix \ref{squeezing}) gives
\be
\langle \zeta(t,\vec k) \zeta(t,\vec k') \rangle = (2\pi)^3 \delta(\vec k + \vec k') \frac{1}{18} \frac{f^2}{\mpl^4 H_0^2}\frac{1}{2 k} \frac1{t^2} \;.
\ee

Although the spectrum clearly shows that the Galileon perturbations are irrelevant on large scales, the reader may be puzzled. Indeed, the two independent solutions of the equation of motion are
\be
\label{zetamodes}
\frac{\sin {k t}}{k t} \; , \quad \frac{\cos {k t}}{kt}
\ee
which, in the limit of long wavelength, respectively give as we discussed
\be
\zeta \sim {\rm const}\;,\;\frac1t \;.
\ee
The first solution describes the celebrated
conservation of $\zeta$ on super-horizon scales\footnote{See
  \cite{Cheung:2007sv} for a proof of the conservation of $\zeta$ at
  all orders in perturbation theory, that applies to the Lagrangian
  studied here.}.  However, this solution is irrelevant as
the second mode dominates at late times. On the other hand we saw
that, in the absence of gravity, we have an attractor and we thus
expect to flow to the adiabatic solution $\zeta$ = const. 
To understand why it is not the $\zeta= {\rm const}$ mode that dominates, in the next Section we move to Newtonian gauge where things will (hopefully) clarify. The Section may however be skipped without loss of continuity.

As to the cosmological (ir)relevance of adiabatic perturbations, in Appendix \ref{squeezing} we study the quantum state of each Fourier mode until it eventually comes back into the Hubble radius, calculating the amount of squeezing induced by the cosmological evolution and comparing our case with inflation. We will see that no appreciable squeezing is produced and all Fourier modes are practically in the ground state when relevant for observations. This unambiguously shows that there are no sizable perturbations.

\section{\label{Newtonian}The two adiabatic modes in Newtonian gauge}
To clarify the situation it is better to write things in a way that has a smooth limit when gravity is decoupled, which is clearly not the case in the gauge we are using, where perturbations of the scalar are set to zero. Things are much clearer in Newtonian gauge
\be
ds^2 = -(1+ 2 \Phi) dt^2 + a^2(t) (1- 2 \Psi) dx^2 \;.
\ee
Instead of doing an explicit change of gauge we can equivalently use the Bardeen potentials (for a recent review on cosmological perturbation theory see \cite{Malik:2008im}), i.e.~the gauge invariant combinations which coincide with the scalar perturbations in Newtonian gauge. These are expressed in terms of our variables
as
\begin{eqnarray}
\Phi & = &\delta N + (a^2 \beta)^{\bf\dot{}} \label{Phi} \\ 
\Psi & = & -\zeta - H a^2 \beta \;. \label{Psi}
\end{eqnarray}
Also the perturbation of the scalar field in Newtonian gauge can be written as a gauge invariant combination which in our notation reduces to 
\be
\label{piGI}
\xi = a^2 \beta \;.
\ee
First of all, let us verify that in this gauge one has a smooth
$\mpl \to \infty$ limit \footnote{The mixing with gravity is important at all scales, but it fades away in the $\mpl \to \infty$ limit with fixed $f$ and $H_0$. Notice however that if one uses \eqref{zetaaction} to write the action for $\varphi$ in the spatially flat gauge, $\zeta = - \varphi \cdot H /\dot \pi_0$, the result does not reduce to the one without gravity, eq.~\eqref{phiaction}, sending $\mpl \to \infty$. The reason for this unexpected result is that if we take $\mpl \to \infty$ in a generic gauge, the spacetime does become flat, but we are not guaranteed that it be written in standard coordinates with metric $\eta_{\mu\nu}$. Indeed it is straightforward to check, starting from \eqref{deltaNzeta} and \eqref{psizeta} and doing the change of gauge, that in spatially flat gauge 
\be
\label{ADMflat}
\delta N_{\varphi} =  - \frac{\dot\zeta}{H} - \left(\frac{\zeta}{H}\right)^{\bf{\dot{}}} \qquad
\beta_{\varphi} = \frac{2}{a^2 H} \zeta + \frac{9 \mpl^2 H_0^2 t^2  }{f^2} \frac1{\nabla^2} \dot \zeta \;.
\ee
The limit must be taken while keeping $\varphi$, i.e.~$\zeta/H$, constant;  we see that both $\delta N$ and $\beta$ remain finite in spatially flat gauge when $\mpl \to \infty$: the metric does not become $\eta_{\mu\nu}$. 
This does not happen if one considers a model with a scalar field with a minimal kinetic term: in this case $\delta N$ and $\beta$ go to zero when $\mpl \to \infty$.}.
We have to check that in this limit, keeping
the amplitude of the scalar mode perturbation $\xi$ fixed, the metric
becomes Minkowski, i.e.~$\Phi$ and $\Psi$ go to zero. From
eq.~\eqref{psizeta} one can see that in the  $\mpl \to \infty$ limit
with fixed  $\xi$, which is equivalent (via eq.~\eqref{piGI}) to
fixed $\beta$, $\zeta$ goes to zero. As $H$ also goes to zero, $\Psi$
in eq.~\eqref{Psi} goes to zero. For $\Phi$ the limit is not so
evident as both terms in eq.~\eqref{Phi} remain finite. However one
can check that they cancel in the limit of decoupled gravity, by using
the equation of motion of $\zeta$ derived from the action
\eqref{zetaaction}. As spacetime approaches Minkowski when $\mpl \to
\infty$, also the equation of motion for the scalar will reduce to that in the absence of gravity.

We are now in a position to compare with the homogeneous perturbations studied in Section \ref{background}. Rewriting $\xi$ in terms of $\zeta$ using \eqref{psizeta} and the expressions \eqref{expearly} we have 
\be
\label{hatpi}
\xi = a^2 \beta = \frac{\zeta}{H} + \frac{\dot H}{H^2} \frac{a^2}{\nabla^2} \dot \zeta \;.
\ee
From this we see that, at long wavelengths, the leading solution $\zeta \propto 1/t$ corresponds to $\xi ={\rm const}$---the attractor we found studying homogeneous perturbations. Let us check that the same holds for the metric as well. In the absence of anisotropic stress $\Psi$ equals  $\Phi$ \footnote{Actually this equality is not so straightforward to obtain. If one expresses $\Phi$ as a function of $\zeta$ using \eqref{Phi} and \eqref{hatpi}, one gets
\be
\Phi = - \frac{\dot H}{H^2} \zeta + \left(\frac{\dot H}{H^2} \frac{a^2}{\nabla^2} \dot\zeta\right)^{\bf{\cdot}} \;.
\ee
There is a partial cancellation between the two terms using the equation of motion of $\zeta$ and the term  $\dot H/ H^2 \zeta$ cancels (notice that $\dot H/ H^2 \gg 1$ in the limit we are considering.) To retain terms ${\cal O}(1) \times  \zeta$---which are needed to check whether $\Phi = \Psi$---one should go beyond the approximation we are using. We do not do so, but we use the expression of $\Psi$ in which no cancellation occurs, taking for granted that subleading corrections will enforce $\Phi = \Psi$.}.  For the $\zeta \sim 1/t$ mode, we can neglect the first term of the right-hand side of eq.~\eqref{Psi} so that
\be
\Psi \simeq - H \xi \qquad {\rm with} \quad \xi = {\rm const.} 
\ee
As $\xi$ = const describes the unperturbed solution shifted in time by an amount $\xi$, we see that the metric also describes the unperturbed FRW shifted in time by $\xi$. Indeed $a^2(t  + \xi) \simeq (1 + 2 H \xi) a^2$ (\footnote{In order to cast the metric in FRW form, one should also do a redefinition of the time coordinate. It is easy to check that the resulting effect on the metric is suppressed by ${\cal O}(H t)$ with respect to $\Psi$.}). 

What is quite confusing is that we are used in standard inflation
to identifying the existence of an attractor with the $\zeta = {\rm const}$ mode's being the leading one at late times. Why does not this happen here? To
understand this different behaviour, it is useful to follow \cite{weinberg} and study the most generic adiabatic mode in Newtonian gauge. One can do so by considering the residual gauge freedom of Newtonian gauge at $k=0 \,$:
\be
t \to t + \epsilon(t) \qquad x^i \to x^i (1- \lambda) \;,
\ee 
with constant $\lambda$.
Under these transformations the potentials transform as
\be
\Psi \to \Psi + H \epsilon - \lambda \qquad \Phi \to \Phi - \dot\epsilon \;.
\ee
These are just gauge modes. To become the $k \to 0$ limit of physical
solutions, they must satisfy Einstein equations at infinitesimal but non-vanishing $k$. These will set $\Phi = \Psi$, assuming
we have no anisotropic stress. Using this, one can find
the most generic adiabatic mode solving for $\epsilon(t)$ 
\be
\label{epsilonW}
\epsilon(t) = \frac{\lambda}{a(t)} \int^t_0 a(t') dt' + \frac{c}{a(t)}\;.
\ee
A generic adiabatic mode is thus fixed by two constants: $\lambda$ and
$c$. Usually one neglects $c$ as $\epsilon$ is dominated by the
integral, which grows as $t$ for any power-law expansion of the form
$a(t) \propto t^{2(1+w)/3}$ with $1+w>0$. This is the standard
adiabatic mode. Notice that the constant $\lambda$ is the value
of $\zeta$ for the constant solution \cite{weinberg}: indeed it parameterizes a rescaling of the spatial
coordinates. For a constant $w$, one obtains a constant value for
$\Phi$
\be
\Phi = \Psi = - \frac{1+w}{\frac53 + w} \lambda \;,
\ee
which is the standard relation between the Newtonian potential and the 
conserved quantity $\zeta$. 
Our situation, however, is quite different. Taking $a \simeq 1$, we see that
the first term in eq.~\eqref{epsilonW} still goes as $t$, but now this
implies that it becomes subdominant as $t$ gets close to zero. The dominant adiabatic
mode is now given by $\epsilon = c/a$, which in our
approximation amounts to $\epsilon \sim c$, corresponding to a constant time-shift. This makes perfect
sense: as the model we are studying has a smooth $\mpl \to \infty$
limit, we expect it to be dominated by an adiabatic mode that reduces to a
constant time shift, which is what we get in the absence of gravity. 

It is immediate to work out the precise relation between the coefficients $\lambda$ and $c$, describing the general long wavelength solution in Newtonian gauge and the two modes of $\zeta$, eq.~\eqref{zetamodes}. We can relate $\Psi$ with $\zeta$ using \eqref{Psi} and \eqref{hatpi}: we see, as already discussed, that the $\zeta \propto \cos{kt}/t$ corresponds to a constant $\epsilon$ and has $\lambda =0$. On the other hand the mode $\zeta = A \sin{kt}/kt$ gives $\epsilon \simeq A t$ with $c=0$ and $\lambda =A$: as expected $\lambda$ corresponds to the amplitude of the constant $\zeta$ mode.

This adiabatic mode enjoys similar properties as the standard $\zeta
= {\rm const}$ one. Independently of any details about the future evolution,
$\epsilon \propto 1/a$ remains a solution. This implies that this mode
is completely irrelevant for observations, as it quickly decays away
in the standard FRW evolution.

To complete the picture, in Appendix \ref{details} we reproduce in Newtonian gauge the results for homogeneous perturbations of Section \ref{background}.




\section{\label{fake}A second scalar in the fake de Sitter}
The conclusion of the previous Sections is that the fluctuations of the $\pi$ scalar are cosmologically irrelevant. We are therefore forced to look for an alternative mechanism to give rise to the observed scale invariant spectrum of primordial perturbations. We do not have to try very hard as the model itself naturally suggests one. The fictitious metric 
\be
g_{\mu\nu}^{(\pi)} \equiv e^{2 \pi(x)} \eta_{\mu\nu} \;,
\ee
with $\pi$ following the unperturbed solution \eqref{pidesitter}, describes de Sitter space. Notice that any coupling  of additional degrees of freedom with $\pi$ will have to go through the metric above to preserve the conformal invariance of the theory and, for de Sitter, any tensor constructed with the metric is proportional to the metric itself. This means that a second scalar $\sigma$ coupled to $\pi$ will behave as in de Sitter space. If $\sigma$ is massless---which can be ensured by a shift symmetry $\sigma \to \sigma + {\rm const}$---its spectrum will be scale invariant,
\be
\langle \sigma(\vec k) \sigma(\vec k') \rangle = (2\pi)^3 \delta(\vec k + \vec k') \frac{H_0^2}{2 k^3} \;,
\ee
while the inclusion of a small mass term will tilt the spectrum either way. It is straightforward to check that the corrections coming from the evolution of the ``real'' metric $g_{\mu\nu}$ are exponentially small, for modes of cosmological interest. This means that a massless $\sigma$ will acquire an {\em exactly} scale invariant spectrum, which is still marginally compatible with the data \cite{Pandolfi:2010dz}. We stress that gravity breaks the conformal symmetry of the Galileon Lagrangian, and this may induce small corrections to the scale invariant spectrum above.

The conversion of $\sigma$ fluctuations into adiabatic ones can happen through one of the mechanisms that have been studied at length for inflation: $\sigma$ may change the way $\pi$ reheats the Universe \cite{Dvali:2003em,Dvali:2003ar}, or become relevant at a later epoch \cite{Lyth:2001nq}.  As the conversion mechanism is model dependent, unfortunately we cannot infer the value of $H_0$ from the data.  However, the experimental signatures are the typical ones for a ``second field" mechanism: large, local non-Gaussianities and, possibly, isocurvature perturbations. 

As it is common to most  alternatives to inflation, the gravitational wave spectrum is very blue in our model, and unobservable. Gravity waves are just sensitive to what the ``real" metric is doing; given the rapid increase of $H$ we will have a very blue spectrum of tensor modes. Indeed given that $a$ can be approximated as a constant while $H$ blows up, each mode gets frozen---$k/a \sim H$---at an amplitude of order of the Minkowski quantum uncertainty. Therefore we have a  spectrum $\propto k^{-1}$, very suppressed on cosmological scales.

\section{\label{super}Faster than light}

We finally analyze the worrisome feature of our scenario: superluminality. It was shown in \cite{Adams:2006sv, Nicolis:2009qm, Nicolis:2008in} that for DGP- and Galileon-like theories superluminal excitations may be generically expected about non-trivial solutions. Let's briefly review the general arguments of \cite{Nicolis:2009qm} and check whether they apply to our case as well.

Consider our model (\ref{minimal}) in the absence of gravity ($M_{\rm Pl} \to \infty$, $g_{\mu\nu} \to \eta_{\mu\nu}$) and about the trivial configuration with $\pi = 0$ (\footnote{As we discussed, fluctuations around $\pi =0$ are ghost-like for the action \eqref{minimal}, but this sign can be flipped starting from a more general Galileon Lagrangian and preserving the NEC violating solution we are interested in. Let us assume thus that the sign is the healthy one: the discussion below remains unaltered.}). $\pi$ excitations are of course exactly luminal, for Lorentz invariance is unbroken. Now turn on localized sources so as to create a weak stationary $\pi_0 (\vec x)$ field. By `weak' we mean that $\pi$'s self-interactions are unimportant to determine the field configuration, that is the solution obeys
\be \label{Laplace}
\nabla^2 \pi_0 \simeq 0 \; ,
\ee
outside the sources. The quadratic Lagrangian for small perturbations $\delta \pi$ about this solution is
\be
\delta_2 {\cal L} = f^2 \,  G^{\mu\nu} \, \di_\mu \delta\pi \,\di_\nu \delta \pi + \dots ,
\ee
where the (inverse) effective metric is
\be
G^{\mu\nu} = \eta^{\mu\nu} \Big(1+ \frac{4}{3 H_0^2} \Box \pi_0 \Big) - \frac{4}{3} \frac{1}{H_0^2} \di^\mu \di^\nu \pi_0 \; , \label{effective_inverse}
\ee
and the dots stand for non-derivative terms for $\delta \pi$, as well as for corrections that are at least quadratic in the background field $\pi_0$ and derivatives thereof. The causal structure is determined by the highest derivative terms in the quadratic Lagrangian---therefore non-derivative ones like mass terms are irrelevant for our discussion. Also irrelevant is the correction proportional to $\eta_{\mu\nu}$ in (\ref{effective_inverse})---at the order in $\pi
_0$ we are keeping, it is  just an overall conformal factor, which does not affect the light-cone aperture. In conclusion, the propagation of $\delta \pi$ is constrained by the light-cone of
\be \label{effective_metric}
G_{\mu\nu} \simeq \eta_{\mu\nu}  + \frac43 \frac{1}{H_0^2} \di_\mu \di_\nu \pi_0 \; .
\ee
Because of (\ref{Laplace}), this is narrower than the Minkowski light-cone in some directions, but wider in others \cite{Nicolis:2009qm}.
Notice that this conclusion only relies on the presence of the Galileon cubic interaction $(\di \pi)^2 \Box \pi$, regardless of its sign---which in fact can be changed by redefining $\pi \to - \pi$.

At the classical level there is no way out---generic weak-field solutions admit superluminal excitations.
At the quantum one however, we have to make sure that the effect be measurable within the effective theory. Roughly speaking, this is the case if signals of frequency lower than the UV cutoff can gain an order one phase-shift  over exactly luminal signals of the same frequency. Suppose $\di \di \pi_0$ in (\ref{effective_metric}) can be approximated as constant over some distance $L$, and let's make a $\delta \pi$ signal travel such a distance. In order for our weak-field approximation to be valid throughout this region we need
\be \label{weak_field}
\di \di \pi_0 \ll H_0^2 \; , \qquad  \di \pi_0 (L) \sim\di \di \pi_0 \times L \ll H_0 \; ,\qquad  \pi_0 (L) \sim  \di \di \pi_0 \times L^2 \ll 1  \; .
\ee
The superluminal shift in the velocity is of order $\delta c \sim \di \di \pi_0 / H_0^2 $, corresponding to an overall phase-shift for $\delta \pi$
\be
\delta {\rm phase} \sim \delta c \, \omega  L \sim \frac{\di \di \pi_0 \, L}{ H_0}\,\frac{\omega}{H_0} \; ,
\ee
where $\omega$ is the signal's frequency.
The phase shift is much smaller than $\omega/H_0$ in the weak field regime---see eq.~(\ref{weak_field})---and becomes of order $\omega/H_0$ if we stretch the linear approximation to the limit. This means that the source of superluminality we identified (there may be others) is ineffective if we declare that the effective theory breaks down at energies/frequencies of order $H_0$, so that the would-be superluminal effect is not measurable consistently within the EFT. Notice that $H_0$ is well below the strong-coupling scale of the theory \cite{Nicolis:2009qm}.

The above discussion applies to small deformations of the trivial $\pi=0$ background. But what about our cosmological solution? As we will see, essentially the same conclusion applies. However the scale $H_0$ will be replaced by $1/t$---not surprisingly given our solution's scale-invariance. Consider first our solution in the absence of gravity, eq.~(\ref{pidesitter}). Given the high-degree of symmetry of the original Lagrangian (SO(4,2)) as well as of the solution (SO(4,1)), small fluctuations about this configuration are exactly luminal  \cite{Nicolis:2009qm}. We now want to run the same argument as above---turning on a weak-field deformation of this background and studying its small excitations---but for simplicity we want to do so in a patch small enough so that we can approximate the ``fake'' de Sitter metric $e^{2 \pi_{\rm dS}} \eta_{\mu\nu}$ as flat. Given the homogeneity of de Sitter space, all points are equivalent, and we can choose for instance $\vec x = 0$, $t = -H_0^{-1}$ as the center of our patch. We can then perform a special conformal transformation combined with a translation---recall that our Lagrangian is conformally invariant---to rewrite the de Sitter background as
\cite{Fubini:1976jm, Nicolis:2008in}
\be
e^{2 \pi_{\rm dS}} = \frac{1}{1+\frac14 H_0^2 \, x^\mu x_\mu} \; ,
\ee
where now $x^\mu$ is measured from our new origin
\footnote{Notice that performing conformal transformations does not perturb the causal structure, even though they do not commute with the Lorentz group, because they only affect the metric by a conformal factor.}.
 This rewriting of the solution makes it immediate to carry out our analysis in a small patch centered at the origin. First of all, the de Sitter conformal factor reduces to
\be
e^{2 \pi_{\rm dS}} \simeq 1 + {2 \pi_{\rm dS}}  \simeq 1 - \sfrac14 H_0^2 \, x^\mu x_\mu \; .
\ee
Second, at lowest order in $H_0^2 x^2$, the full non-linear dynamics of perturbations about the $\pi_{\rm dS}$ background are given by a Galileon Lagrangian  whose coefficients are of the same order as the original ones \cite{Nicolis:2008in}. In other words, expanding a generic Galileon Lagrangian about a $\pi \propto x^\mu x_\mu$ solution yields another Galileon Lagrangian with similar coefficients. As a consequence, as long as we restrict to distances from the origin smaller than $H_0^{-1}$, our analysis above applies unaltered. We thus have superluminal excitations, which do not really have a chance of yielding measurable superluminal effects if the effective theory breaks down at frequencies of order $H_0$, or below.

However here the appearance of $H_0$ stems uniquely from our choosing to expand about $t=H_0^{-1}$. This is made manifest by working with the field
\be
\phi = H_0 e^\pi \; ,
\ee
in terms of which the Lagrangian takes the form \cite{Nicolis:2009qm}
\be
{\cal L} = \frac{f^2}{H_0^2} \phi^4 F\Big( \frac{\di \phi}{\phi^2}, \frac{\di \di \phi}{\phi^3} \Big) \; ,
\ee
where $F$ is a polynomial with order-one coefficients.
The overall dimensionless factor $f^2/H_0^2$  has no effect at the level of classical equations of motion. Then the only scale present in the Lagrangian is the local value of $\phi$. Our de Sitter solution (in the original coordinates) corresponds to 
\be
\phi_{\rm dS} = -\frac{1}{t} \; .
\ee
So, the fact that to avoid superluminality in a neighborhood of $t = H_0^{-1}$ we have to impose a frequency cutoff of order $H_0$, implies that to avoid superluminality about a generic $t$ the UV cutoff has to be $1/|t|$. Notice that cosmological perturbations of the Galileon field freeze-out at frequencies precisely of order $1/|t|$. 
Therefore, if we decide to ban superluminality from our effective theory, we also lose predictivity for cosmological observables.
The other possibility---swallowing the presence of measurable superluminal effects in a Lorentz-invariant effective theory---does not necessarily lead to inconsistencies. As long as the effective theory is free of closed time-like curves (which for our model has not been proven nor disproven yet), there are no pathologies from the low-energy viewpoint (see for example \cite{Babichev:2007dw} for an optimistic point of view). However, physically measurable superluminality certainly implies that the effective theory at hand cannot arise as the low-energy limit of a microscopic theory with the standard relativistic causal structure, such as a renormalizable Lorentz-invariant QFT for instance \cite{Adams:2006sv}.

The introduction of dynamical gravity perturbs the above analysis, and to some extent its conclusions too. To begin with, our cosmological solution (\ref{pidesitter}) gets modified: slightly at early times (eq.~(\ref{earlytimepi})),  drastically at late ones (eq.~(\ref{polephase})). As a consequence its de Sitter symmetry is gone, and small $\delta \pi$ perturbations are no more exactly luminal, even in the absence of the sources we needed to run the above arguments. Second, given the peculiar structure of the Galilean self-interactions, the mixing of $\delta \pi$ with scalar gravitational perturbations is relevant at all scales (see a related discussion in \cite{markus}), and cannot be ignored even when studying sub-horizon perturbations.
We thus have to use eq.~(\ref{zeta_action}), which is the quadratic action for the propagating scalar mode about the FRW background, taking into account gravitational corrections. The propagation speed for short-wavelength excitations as measured by a comoving observer in terms of the background FRW metric is 
\be
c^2_\zeta = \frac{B(t)}{A(t)} \; .
\ee
In other words, if $c^2_\zeta$ thus defined is larger than one $\zeta$ excitations exit the FRW light-cone. Plugging the approximate solutions we found in  sect.~\ref{background} into eqs.~(\ref{A(t)}, \ref{B(t)}), we get
\bea
 t \to -\infty : & \quad & c^2_\zeta \simeq 1- \frac{32}{9} \frac{f^2}{\mpl^2} \frac{1}{H_0^2 t^2 }    \\
 t \to t_0 : & \quad & c^2_\zeta \to 0  \; .
\eea
In both regimes the correction to the propagation speed is {\em sub}-luminal---extremely so at late times
\footnote{The vanishing of $c^2_\zeta$ for $t$ approaching $t_0$ signals that, in such a limit, higher-derivative corrections to the perturbations' gradient energy cannot (and should not) be neglected, pretty much like for the ghost condensate \cite{ArkaniHamed:2003uy}.}.
This relaxes our conclusions above somewhat: since the cosmological background introduces a subluminal offset into the excitations' speed, we now need perhaps small, but finite perturbations to overturn this offset and make excitations about a new background superluminal.
Although this is certainly more welcome than the opposite result, in practice it is not very helpful: at very early times the gravitational correction to the propagation speed goes to zero---like all other gravitational effects in our model. This means that, at least at early times, we have to live with superluminality, or give up the model.

\section{\label{conclusions}Conclusions and Outlook}

We are putting forward a model for our Universe's early cosmology that departs strikingly from the conventional inflationary picture. Schematically: there is no Big Bang in our past; spacetime is flat  at $t \to -\infty$; related to this, the Universe is initially  devoid of any form of energy; energy and the associated Hubble expansion get created by a NEC-violating sector.

The main virtue of the model lies not in its sheer radicalness---we are certainly not the first authors to come up with ``phantom''-like equations of state---but in its being able to associate such a radicalness with an healthy effective field theory coupled to gravity, and in the consequent robustness of the scenario. Our theory---the Galileon or more precisely its conformally invariant generalization---is well-behaved classically as well as quantum mechanically, even for strongly non-linear background solutions. The structure of the Lagrangian is protected by symmetries.  More relevant for us, the system
 retains stability even when the stress-energy tensor violates the NEC \cite{Nicolis:2009qm}. Partially as a consequence of this, the cosmological solution we outlined is an attractor---the universe wants to follow it even when initially displaced from it---which implies that our model solves the horizon and flatness problems as well as standard inflation does. Remarkably, expansion is not put in as an initial condition but follows from generic initial conditions, including contracting ones---within a bounded  basin of attraction of course.
As to density perturbations, in its minimal incarnation our model does {\em not} produce sizable adiabatic perturbations on cosmological scales. However the symmetry structure is so constraining that postulating the existence of extra light scalars {\em unavoidably} yields nearly scale-invariant spectra for them---which can later be converted into adiabatic perturbations via any of the standard conversion mechanisms available on the market. The downside is that, like for standard multi-field inflationary models, predictions are more model-dependent than for single-field slow-roll inflation, although the presence of a sizeable local Non-Gaussianity is rather robust.

The only reservation we have about welcoming our model as a compelling alternative to inflation concerns superluminality.
The generic presence of superluminal excitations about non-trivial solutions indicates that our model cannot arise as the low-energy limit of a standard relativistic UV-complete theory, like e.g.~a renormalizable Lorentz-invariant QFT. Depending on one's personal taste and attitude, reactions to giving up Lorentz invariance in the UV may range from disgust to excitement.
Minimally, it is fair to say that it makes us depart from known territory, especially when gravity is involved. We therefore feel that it deserves special care. It is interesting to note that superluminality in our model is tied to the presence of the DGP-like interaction $(\di \pi)^2 \Box \pi$, which in turn is forced upon us by demanding that scattering amplitudes obey standard properties of $S$-matrix theory \cite{Nicolis:2009qm}. However it is possible, in principle, that suitable deformations of the theory exist where such an interaction is absent and where superluminality is gone as well.\footnote{Given the non-renormalization theorem of \cite{LPR}, such a ``tuning'' would be preserved by quantum corrections.}
Of course we would like to maintain the nice features of our cosmological scenario---among which the near deSitter invariance of the solution. So, instead of considering a conformal completion of the Galileon theory, we may consider a different symmetry group containing the de Sitter one as a subgroup and that reduces to the galileon symmetry group in the appropriate limit. An obvious choice is the 5D Poincar\'e group ISO(4,1) \cite{Nicolis:2008in, Nicolis:2009qm, dRT}.
This possibility certainly deserves further study.


\section*{Ackowledgements}
It is a pleasure to thank Riccardo Rattazzi for collaboration in the early stages of this project and Niayesh Afshordi for useful discussions.

\begin{appendix}
\section{\label{details}Details on adiabatic perturbations}
This Appendix complements the study of adiabatic perturbations of Sections \ref{perturbations} and \ref{Newtonian}. 

{\em Unitary gauge action}. To deduce eq.~\eqref{piADM}, we use that in unitary gauge the various terms in the action read
\bea
\sqrt{-g} \, e^{2\pi} (\di \pi)^2 & = & - e^{2 \pi_0}\dot \pi_0^2  \, \sqrt{g_3} \frac{1}{N}\\
\sqrt{-g} \, (\di \pi)^4 & =  & \dot \pi_0^4 \, \sqrt{g_3} \frac{1}{N^3} \\ 
\sqrt{-g} \,  \Box \pi (\di \pi)^2 & = & -2 \dot \pi_0^2 \ddot \pi_0 \, \sqrt{g_3} \frac{1}{N^3} + \dot \pi_0^3  \, \sqrt{g_3} \frac{1}{N} \left[ N^i \di_i \frac{1}{N^2} - \di_t \frac{1}{N^2} \right]  \label{dgp_term}\; ,
\eea
where $g_3 \equiv \det g_{ij}$, we used that $ \sqrt{-g} = N \sqrt{g_3}$, and in the last line we integrated by parts. Also we made use of $g^{0i} = N^i / N^2$, where and henceforth spatial indices are raised and lowered with the spatial metric $g_{ij}$. We can rewrite the terms in bracket in terms of the extrinsic curvature of costant-$t$ hypersurfaces,
\be
K_{ij} \equiv \frac{1}{2N} \big[ \di_t g_{ij} - \nabla_i N_j - \nabla_j N_i \big] \; .
\ee 
Indeed after straightforward manipulations and integrating by parts we get
\be
\dot \pi_0^3  \, \sqrt{g_3} \frac{1}{N} \left[ N^i \di_i \frac{1}{N^2} - \di_t \frac{1}{N^2} \right] = 2  \dot \pi_0^2 \ddot \pi_0 \, \sqrt{g_3} \frac{1}{N^3} + \sfrac23  \dot \pi_0^3  \, \sqrt{g_3} \frac{1}{N^2}  K^i {}_i  \; ,
\ee
where we used that $\di_t \sqrt{g_3} = \frac12 \sqrt{g_3} \, g^{ij} \, \di_t g_{ij}$. The first piece cancels exactly the first term in eq.~(\ref{dgp_term}) and we are left with eq.~\eqref{piADM} (\footnote{As $\ddot\pi \dot\pi^2$ is a total derivative, each term in $(\partial\pi)^2\Box\pi$ contains at least two spatial derivatives. That's why it is not surprising that it can be written solely in terms of an operator containing the extrinsic curvature $K$, which contains two spatial derivatives on $\pi$.}).

{\em Unitary action in the standard form}. To cast eq.~\eqref{piADM} in the form \eqref{ADMaction} we can expand the third term of  \eqref{piADM} as $1/N^4 = 2/N^2 -1 + 4 \delta N^2 + \ldots$ The second term can be rewritten as
\be
\frac{1}{N^3} K^i_i = \delta\frac{1}{N^3}\delta K_i^i + K^i_i +3 H \frac{1}{N^3} -3 H \;.
\ee
Notice that $K^i_i$ appears as an additional tadpole term besides the ones in eq.~\eqref{ADMaction}. However one can get rid of it using the identity \cite{Cheung:2007st}\footnote{Notice there is a sign error in the last term of eq.~(80) of \cite{Cheung:2007st}.} 
\be
\int d^4 x \sqrt{-g} \, f(t) K^\mu_\mu = \int d^4 x \sqrt{-g} \, f(t) \nabla_\mu n^\mu =  -\int d^4 x \sqrt{-g} \,\partial_\mu f(t) n^\mu= - \int d^4 x \sqrt{-g} \,\dot f (t) \frac{1}{N} \;.
\ee
In this way one can write the action in the form \eqref{ADMaction} and check that the coefficients of the tadpole terms can indeed be written in terms of $H$ and $\dot H$ using the expression of the $\pi$ stress-energy tensor.

{\em Constraint equations}. The explicit form of the constraint equations is
\bea
\mpl^2 \bigg[R_3 -  \frac1{N^2}\big( E_{ij}E^{ij} - E^i {}_i {}^2 \big) + \frac{2}{N^2} \dot H
-2 \big( 3H^2 + \dot H\big) \bigg] + 2 M^4 \delta N 
- 2 \hat M^3 \delta E^i {}_i =   0 && \\
\nabla_i \bigg[ \mpl^2 \frac1N \big(E^i{}_j - \delta^i_j \, E^k {}_k \big) -  \delta^i_j \, \hat M^3  \delta N\bigg]  =   0 &&
\eea
whose solution at linear order gives \eqref{deltaNzeta} and \eqref{psizeta}.

{\em Homogeneous perturbations in Newtonian gauge}. Let us check that in Newtonian gauge we find the same two regimes of perturbations we found in Section \ref{background}: one when the perturbation dominates the energy density (this is equivalent to decoupling gravity as the energy density of the background vanishes when $\mpl \to \infty$) and one when perturbations are small and only perturb the background energy density. The first regime is obtained simply  by sending $\mpl \to \infty$. In eq.~\eqref{hatpi} the leading solution in $\zeta$ gives, as we saw, the $\xi = {\rm  const}$ mode, while the solution $\zeta \propto \sin{kt}/kt$ gives the decaying solution $\xi \sim t^5$, as it can be easily verified. Actually this identification holds not only on large wavelengths, but it is exact in the limit $\mpl \to \infty$: the second mode of $\zeta$ in eq.~\eqref{zetamodes} matches with the solution of eq.~\eqref{phiaction} that is constant at small $k$, while the first one matches with the $\xi$ mode that as $t^5$ at small $k$, without mixing. It is much trickier to study the regime when perturbations do not dominate the energy density. From Section \ref{background} we expect the decaying mode to give $\xi \sim t^2$ in this case, but as we said the decaying mode of $\zeta$ gives $\xi \sim t^5$. The trick is that one has to be careful about the two limits $\mpl \to \infty$ and $k \to 0$. Indeed, if one keeps higher orders in $1/\mpl^2$ in the action for $\zeta$, eq.~\eqref{zetaaction}, in the relation between $\zeta$ and $\xi$, eq.~\eqref{hatpi}, and takes also into account the change in the time variable to compare with the FRW solutions of Section \ref{background}, the decaying mode goes as
\be
\xi \propto \frac{H_0^2 t^5 k^2}{5 f^2} \mpl^2 + \left(-\frac{13 \pi}{5 k} + \frac{13}{15} \pi t^2 k\right) + \ldots
\ee
where the dots stand for terms of higher order in $1/\mpl^2$ and $k$. We see that the $\xi \sim t^5$ solution dominates if one sends $\mpl \to \infty$ at fixed $k$. On the other hand sending $k \to 0$ at fixed $\mpl$ we have a constant term, which describes a mixing with the dominant $\xi = {\rm const}$ mode and the $t^2$ behaviour we were looking for.

\section{\label{squeezing} Squeezing and absence thereof}
As we discussed, the perturbations of the Galileon field $\pi$ have quite peculiar properties, rather different from the standard inflationary scenario. The best way to pin down the status of these perturbations is to determine the quantum state of each Fourier mode, when it comes back into the Hubble radius, for example during a radiation dominated phase. Indeed, in the linear approximation, the field is just a collection of harmonic oscillators, each with a time dependent Lagrangian, as the background we are perturbing around is time dependent.
This causes each Fourier mode to be in a squeezed state, when it gets back into the Hubble radius. The amount and direction of squeezing uniquely fix the state of the perturbation and also tell us whether a classical interpretation in terms of classical stochastic variables is possible. 

In order to follow the evolution of each Fourier mode until the it comes back into the Hubble radius, it is useful to have a unified description in which perturbations around a homogeneous, isotropic Universe are always described in each phase of the evolution of the Universe by an action for the same scalar variable. We will do so using $\zeta$ as such a variable. In this Appendix we start calculating the squeezing status of perturbations in the case of inflation and then compare it to our Galileon case. Of course the case of inflation is quite well known, but the way it is presented here is, to our knowledge, new. 

The action for $\zeta$ during inflation is of the form \cite{maldacena}
\be
\label{inflaction}
S= \mpl^2 \int d^4 x \;a^3 \epsilon \left[\dot\zeta^2- \frac{1}{a^2} (\nabla\zeta)^2\right] \;,
\ee
where $\epsilon = \dot\phi^2/(2 H^2 \mpl^2)$ can be taken as a constant at leading order in slow-roll. The field $\zeta$ can be decomposed in terms of annihilation and creation operators as
\be
\label{zetaaa}
\zeta(t,\vec x) = \int \frac{d^3k}{(2\pi)^3} \left(\zeta_{\vec k}^{\rm cl}(t) a_{\vec k} + \zeta_{\vec k}^{{\rm cl}*}(t) a^\dagger_{\vec k}\right)\;.
\ee
As the field satisfies the equations of motion, the functions $\zeta_{\vec k}^{\rm cl}(t)$ are solutions of the equations of motion that at very early times, when the mode is much shorter than the Hubble radius, reduce to the Minkowski form: 
\be
\zeta_{\vec k}^{\rm cl}(t) = \frac{1}{\sqrt{2 \epsilon} \mpl}  \cdot \frac{H}{\sqrt{2k^3}}\left(1 - i \frac{k}{a H}\right) e^{i \frac{k}{a H} + i \vec k \vec x} \;.
\ee
Let us define a scalar product between classical solutions of the field equation
\be
\label{scalar}
\langle \zeta_1^{\rm cl} ; \zeta_2^{\rm cl} \rangle \equiv - i \int d^3x \left( \zeta_1^{\rm cl} \Pi_2^{\rm cl *} -  \zeta_2^{\rm cl *} \Pi_1^{\rm cl}\right) \;,
\ee
where $\Pi^{\rm cl}$ is the momentum conjugate to $\zeta$. For the action \eqref{inflaction}, $\Pi^{\rm cl} = 2 \mpl a^3 \epsilon \dot\zeta^{\rm cl}$.  It is important to notice that this scalar product is time independent as a consequence of the equations of motion. The solutions $\zeta^{\rm cl}$ are normalized as
\be
\begin{split}
\label{norms}
\langle \zeta_{\vec k}^{\rm cl} ; \zeta_{\vec k'}^{\rm cl} \rangle & = (2 \pi)^3 \delta(\vec k - \vec k') \\
\langle \zeta_{\vec k}^{\rm cl *} ; \zeta_{\vec k'}^{\rm cl *} \rangle & = - (2 \pi)^3 \delta(\vec k - \vec k') \\
\langle \zeta_{\vec k}^{\rm cl} ; \zeta_{\vec k'}^{\rm cl *} \rangle & = 0 \;.
\end{split}
\ee

The action describing scalar perturbations during a phase dominated by a barotropic fluid with $p = w \rho$ is given by \cite{Boubekeur:2008kn}
\be
S= \mpl^2 \int d^4 x \;a^3 \frac{3(1+w)}{2w} \left[\dot\zeta^2- \frac{w}{a^2} (\nabla\zeta)^2\right] \;.
\ee
Let us concentrate for example on a period of radiation dominance, $w=1/3$, which gives an evolution $a \propto t^{1/2}$, $H = 1/(2 t)$. One can still perform an expansion analogous to \eqref{zetaaa}, 
\be
\label{zetaaa2}
\zeta(t,\vec x) = \int \frac{d^3k}{(2\pi)^3} \left(\tilde\zeta_{\vec k}^{\rm cl}(t) \tilde a_{\vec k} + \tilde\zeta_{\vec k}^{{\rm cl}*}(t) \tilde a^\dagger_{\vec k}\right)\;.
\ee
Now the appropriate solutions of the equation of motion are given by 
\be
\tilde \zeta_{\vec k}^{\rm cl}(t) = \frac{1}{\sqrt{12} \mpl}\frac{1}{\sqrt{2k/\sqrt{3}}} \cdot \frac{i}{a}  e^{- i \frac{1}{\sqrt{3}}\frac{k}{a H} + i \vec k \vec x} \;.
\ee
These functions reduce to the Minkowski result in the limit in which the modes are well within the Hubble radius and they are normalized in the same way as in \eqref{norms}. The choice of phase of these solutions is done for later convenience.

Equating the two expansions \eqref{zetaaa} and \eqref{zetaaa2} for each Fourier mode we get
\be
\label{2Fourier}
\zeta_{\vec k}^{\rm cl} a_{\vec k} + \zeta_{-\vec k}^{{\rm cl}*} a^\dagger_{-\vec k} = \tilde\zeta_{\vec k}^{\rm cl} \tilde a_{\vec k} + \tilde\zeta_{-\vec k}^{{\rm cl}*} \tilde a^\dagger_{-\vec k}
\ee
The two set of modes are related by
\be
\label{modesrel}
\tilde \zeta_{\vec k}^{\rm cl} = \alpha_{\vec k} \zeta_{\vec k}^{\rm cl} + \beta_{\vec k} \zeta_{-\vec k}^{\rm cl *}
\ee
where the coefficients can be generically calculated using the scalar product\footnote{Notice that the scalar product \eqref{scalar} is well defined at any time, and time independent. Of course the explicit expression for the momentum $\Pi$ in terms of $\dot \zeta$ depends on the action for $\zeta$ valid at any given moment.} 
\be
\alpha_{\vec k} = \langle \tilde \zeta_{\vec k}^{\rm cl} ; \zeta_{\vec k}^{\rm cl}\rangle \qquad \beta_{\vec k} = -\langle \tilde \zeta_{\vec k}^{\rm cl} ; \zeta_{-\vec k}^{\rm cl *}\rangle \; ,
\ee
but at small $k$ we can use a quicker method---see below.
Plugging eq.~\eqref{modesrel} into \eqref{2Fourier} gives the relation among the two sets of creation and annihilation operators
\be
a_{\vec k} = \alpha_{\vec k} \tilde a_{\vec k} +  \beta^*_{\vec k} \tilde a^\dagger_{-\vec k} \;.
\ee
We assume to be in the vacuum at the beginning, i.e.~in a state which is annihilated by $a_{\vec k}$. This state does not evolve in time as we are putting the time evolution in the operators. In terms of the ``radiation dominance" operators $\tilde a_{\vec k}$, the state of each harmonic oscillator is annihilated by a linear combination of $\tilde a_{\vec k}$ and $\tilde a^\dagger_{\vec k}$: it is a squeezed state.
Taking the norm of both sides of eq.~\eqref{modesrel} we derive that the coefficients $\alpha$ and $\beta$ must satisfy
\be
\label{bogolubov}
|\alpha_{\vec k}|^2-|\beta_{\vec k}|^2 =1
\ee
which is equivalent to the condition that both sets of creation and annihilation operators satisfy the standard commutation rules. 

The simplest way to relate the modes $\zeta^{\rm cl}$ and $\tilde\zeta^{\rm cl}$, is noticing that their real parts become constant in the long wavelength limit, while their imaginary parts have constant conjugate momenta in the same limit. Given the independence or incompatibility of these two conditions---constant field vs.~constant momentum---this implies that real and imaginary parts do not mix in the long wavelength limit, that is
\be
\zeta_{\vec k}^{\rm cl} + \zeta_{\vec k}^{\rm cl *} = A_k (\tilde\zeta_{\vec k}^{\rm cl} + \tilde\zeta_{\vec k}^{\rm cl *})  \; , \qquad \zeta_{\vec k}^{\rm cl} - \zeta_{\vec k}^{\rm cl *} = B_k (\tilde\zeta_{\vec k}^{\rm cl} - \tilde\zeta_{\vec k}^{\rm cl *}) \; , \qquad k \to 0 \;.
\ee
Condition \eqref{bogolubov} implies that the two coefficients $A_k$ and $B_k$ are the inverse of each other: $A_k = B_k^{-1}$. $A_k$ and $B_k$ can be calculated using the explicit expression of the modes: it is enough in the two cases to compare the value of $\zeta$ and $\Pi$ which is approached in the long wavelength limit. One gets
\be
\label{Ainfl}
A_k =  3^{3/4} \sqrt\frac{2}{\epsilon} \frac{a_{\rm rd}^2 H_{\rm rd}H_{\rm infl} }{k^2}  \qquad B_k = \frac{1}{A_k} \;.
\ee
These expressions are time-independent, as they should. It is easy to realize that, neglecting numerical factors, one has
\be
A_k \sim \frac{a_{\rm in}}{a_{\rm out}} \gg 1 \;,
\ee
the ratio between the scale factors when the modes leave the Hubble radius and when they come back in. The uncertainty in the real part of $\zeta$ is huge compared to that we would have in the ``radiation dominance vacuum". Conversely the uncertainty in the imaginary part is very suppressed with respect to the vacuum state. If we neglect this minuscule uncertainty in the imaginary part, assuming that fluctuations in every observable will be dominated by the huge real part's uncertainty, we can treat the quantum field as a classical stochastic variable.

Let us now come back to our model, and carry out exactly the same procedure. Now the $\zeta_{\vec k}^{\rm cl}$ modes must be calculated from the action \eqref{zetaaction} and they are given by
\be
\zeta_{\vec k}^{\rm cl} = \frac{f}{3 \sqrt{2} \mpl^2 H_0}  \frac{1}{\sqrt{2k}} \frac{i}{t} e^{-i k t + i \vec k \vec x} \;, 
\ee
while the radiation dominance modes $\tilde\zeta_{\vec k}^{\rm cl}$ remain, obviously, the same. Again the real part of the modes gives $\zeta$ = const in the long wavelength limit, while the imaginary part has $\Pi$ = const. Matching these constants allows to calculate the coefficients $A$ and $B$
\be
\label{Agali}
A_k = \frac{f}{\mpl H_0} a_{\rm rd}^2 H_{\rm rd} \frac{\sqrt{2}}{3^{1/4}} \qquad B_k = \frac{1}{A_k} \;.
\ee  
Let us assume that the Galileon dominated phase ends when $H \simeq \mpl/f H_0$, i.e.~when the approximation of treating gravity as a perturbation breaks down, and that immediately after we have a radiation dominated phase. Evaluating $A_k$ at the transition between the two phases we get
\be
A_k \sim B_k \sim 1 \;.
\ee
There is no relevant squeezing of the modes, so that during radiation dominance the uncertainties are close to the standard zero point quantum fluctuations. Of course in this case the perturbations are completely negligible and no classical interpretation is possible\footnote{The complete absence of squeezing is accidental and does not occur if we modify the matching between the Galileon dominated regime and the standard decelerating evolution, to take into account the intermediate phase when gravity cannot be treated as a small perturbation to the Galileon dynamics, eqs.~\eqref{polephase} and \eqref{polephase2}. The same will happen if we replace radiation dominance with a different decelerated evolution. However, by comparing \eqref{Agali} with \eqref{Ainfl},  we see that the squeezing parameters are very ``blue" compared with the inflationary result, and therefore always irrelevant on cosmological scales.}.

A sizable generation of perturbations needs a large squeezing. This, in inflation, is closely related to the existence of a dynamical attractor which makes the $\zeta = {\rm const}$ mode dominate. Our model shows that the existence of an attractor does not by itself guarantee a sizable generation of perturbations:  during radiation dominance the modes are essentially in their vacuum state.

\end{appendix}


\end{document}